\documentclass[aip,graphicx]{revtex4-1}
\usepackage{graphicx}
\usepackage{dcolumn}
\usepackage{bm}

\draft 

\begin{document}


\title{Sensing and cooling of a nanomechanical resonator with an electron beam stimulated internal feedback and a capacitive force} 



\author{A. Descombin}
\author{S. Perisanu}
\author{P. Poncharal}
\author{P. Vincent}
\author{S. T. Purcell}
\author{A. Ayari}%
 \email{anthony.ayari"@"univ-lyon1.fr}
\affiliation{%
Univ Lyon, Universit\'e Claude Bernard Lyon 1, CNRS, Institut Lumi\`ere Mati\`ere, F-69622, LYON, France.}



\date{\today}

\begin{abstract}
A model for the cooling properties of a nanocantilever by a free electron beam is presented for a capacitive interaction. The optimal parameters for position sensing and cooling applications are estimated from  previous experimental conditions. In particular, we demonstrate that a purely capacitive force and an electron beam stimulated internal feedback  can lower the temperature of a nanocantilever by several orders of magnitude in striking contrast with the conventional  electrostatic damping regime. We propose a step by step protocol to extract the interdependent parameters of the experiments. This work will aid future developments of ultra sensitive force sensors in electron microscopes.
\end{abstract}


\maketitle 


\section{Introduction}
The recent works on nanoelectromechanical systems (NEMSs) have led to considerable improvements of their mass\cite{chaste2012} and force\cite{moser2013ultrasensitive} sensing limit. Such a progress was the result of a careful control of the coupling strength with the detection probe, yielding to a trade-off between the need to increase the signal from the small amplitude of the NEMS vibration and a limitation of the concomitant increase of the probe perturbations such as heating, additional noise and non-linear effects. To the opposite, a sufficiently strong and efficient perturbation due to the probe can produce a backaction force that induces a cooling of a NEMS down to the quantum regime\cite{teufel2011sideband} and be used for coherent electron-photon conversion\cite{bagci2014optical} and quantum communication\cite{o2010quantum}. Such perturbations can be performed by an external feedback loop\cite{Kawa} with an amplifier and a phase shifter but an interesting feature of NEMS devices is the possibility to form an internal feedback loop\cite{Weig2012,ayari_self_oscillations,barois2014frequency} where the active component is hidden among the intrinsic elements of the device and their coupling.

The probe coupling for position sensing and backaction cooling has been extensively studied in the framework of gravitational waves\cite{braginsky1995quantum} and optomechanics\cite{aspelmeyer2014cavity,PhysRevB.70.245306} where photons are used as a probe of the mechanical vibrations. To the contrary the coupling between an electron beam and a nanoresonator has been little studied and is hardly ever used for NEMS (see ref .\cite{PhysRevB.63.033402,vincent2007driving,yasuda2016oscillation,siria2017electron} and references therein) despite the fact that focused free electron beams have a higher spatial resolution than lasers and are essential for imaging applications. A known drawback of using an electron beam on a nanostructure is the deposition of amorphous carbon\cite{gil2010nanomechanical} due to the dissociation of organic residue present in the SEM chamber. However this contamination can be prevented by in situ plasma cleaning, cryopumping or using a load lock to keep the chamber clean. Another issue is the capacitive force induced by the charging of the NEMS by the electron beam. This force can result in the collapse of the nanostructure if the vibrating part is too close to a counter electrode. Apart from damaging the sample, it is usually considered that a capacitive force cannot lead to cold damping\cite{aspelmeyer2014cavity} without an external feedback\cite{jourdan2007tuning}. An increase of the mechanical damping\cite{jourdan2007tuning,barois2012ohmic}, sometimes called electrostatic damping, arises when a DC voltage is applied to a NEMS cantilever. This damping leaves the temperature of the mechanical mode unchanged in contrast to the usual coupling to a red detuned cavity. From this, it might mistakenly be inferred that the internal feedback loop formed by a NEMS and the capacitive force induced by an electron source will also lead to damping without cooling. In fact the type of applied bias and the spatial dependence of the electrical components play an equally important role to predict the behavior of such systems. For example, it has been shown that a motional resistance can lead to self-oscillations in capacitive NEMSs\cite{barois2013ultra} or hysteresis and memresistive behavior in a carbon field emitter\cite{gorodetskiy2016memristive}. Recently, it was also demonstrated that the capacitive dynamical backaction on a NEMS has an opposite effect if an AC voltage is applied instead of a DC voltage.\cite{eriksson2015nonresonant,PhysRevApplied.6.014012}

In this work, we present a more in-depth analysis of the interaction with a free electron beam compared to our previous work\cite{vincent2007driving}. An analytical model with a purely capacitive backaction force is developed in order to clarify the competition between electrostatic damping and electron stimulated cold damping. In particular, It will be shown that although electrostatic damping and electron stimulated cold damping come from the same capacitive force, retarded by the same mechanism, their respective effect can be dramatically different because in one case the electrical circuit is voltage driven whereas in the other it is current driven. Our analysis, will highlight the fact that a) in the electrostatic damping regime, the mechanical system is submitted to an additional stochastic force originating from the Johnson-Nyquist noise of the resistor in thermal equilibrium with the room temperature bath; b) in the electron stimulated cold damping regime, this stochastic capacitive force comes from the noise of the electron gun and for low current the effective temperature of this source can be lower than room temperature. The optimal conditions for cooling and self-oscillations are estimated based on previous experiments and confirmed by numerical simulations. Experimentally accessible data usually present an interdependency on the different degrees of freedom that makes difficult the estimation of the experimental parameters. Therefore, a thorough experimental protocol is established in order to extract key parameters from data. The performances and the limitations of this free electron beam position sensor is then studied and compared to other electronic sensing techniques in NEMS\cite{mi8040108} such as quantum point contacts\cite{flowers2007intrinsic}, current mixing with a single electron transistor\cite{PhysRevB.95.035410} and field emission\cite{ayari_self_oscillations,Victor2009,lazarus,PhysRevBself}.

\section{analytical model}
\subsection{Mechanical equations}
\begin{figure}
\includegraphics{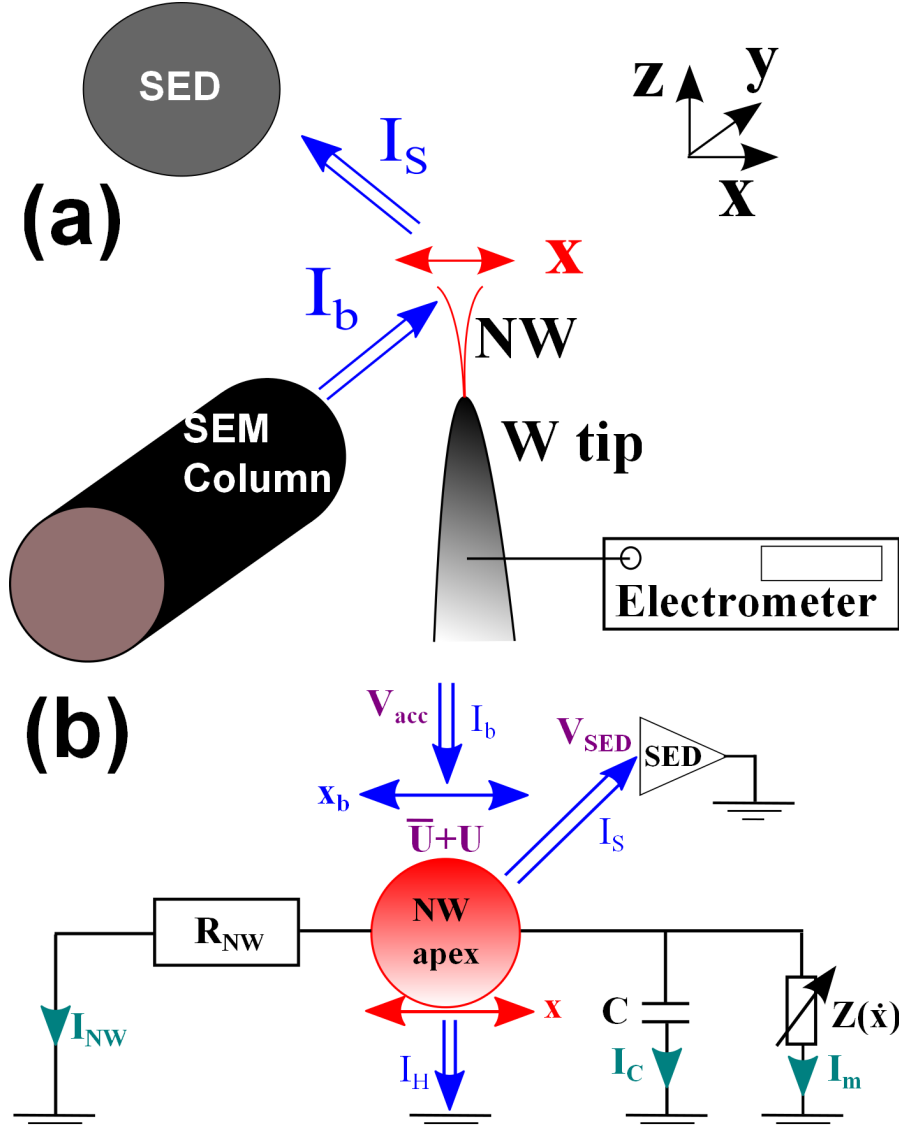}
\caption{\label{figelec} a) Geometry of the experimental set up b) Schematic of the electrical circuit with the different components. The double line arrows represent the free electron currents. The horizontal arrows indicate the direction of motion of the nanowire apex and the focal point of the primary electron beam.}
\end{figure}

We consider a single clamped nanowire (NW) in a scanning electron microscope (SEM) environment. The SEM beam is perpendicular to the resonator and focused at its free end (see figure \ref{figelec} a). The model can be easily extended for the case where the electron spot is at a different position along the nanowire. The nanocantilever can vibrate in the transverse direction along two perpendicular mechanical polarizations. We suppose that one of these polarizations has been pre-aligned perpendicularly to the electron beam. The electron beam interacts with the nanowire and exerts an actuation force $F_e$ on the nanocantilever with both a deterministic and a stochastic component. The mechanical dynamical equation is:
\begin{equation}
\label{act1}
m(\frac{d^2}{dt^2}+\Gamma_0\frac{d}{dt}+\omega_0^2)(\bar{x}+x) = F_e + F_T
\end{equation}
where m is the effective mass, t is the time, $\omega_0/2\pi$ is the resonance frequency, $\Gamma_0$ is the damping, $\bar{x}$ is the averaged displacement perpendicular to both the nanowire and the SEM beam, x is the instantaneous additional displacement, $F_T$ is the stochastic thermal force.

We will limit our analysis to a one dimensional problem and suppose that the two polarizations are uncoupled. There is no mechanical coupling such as observed in ref. \cite{PhysRevB.81.165440} because of the small amplitude of vibration and no coupling due to the electron beam force. When the validity of this last assertion is not satisfied experimentally, this can lead to richer dynamical effects, such as those observed in ref.\cite{gloppe2014bidimensional}. The physical effect, we want to point out here can be integrated into a 2D model but for simplicity we will neglect these couplings. The interaction with electrons can be described by three different backaction forces : a capacitive force, a thermal force stemming from the electron kinetic energy absorbed by the NW and a direct momentum transfer force. In ref\cite{vincent2007driving} we showed that by changing the capacitive environment, we drastically modified the way the nanowire interacts with the electron beam and were capable to reverse the conditions where self-oscillations take place. Such an effect cannot be explained by thermal or momentum transfer forces. Despite apparent system symmetry, the experimental capacitive environment is not symmetric, so that moving the nanowire to one direction will increase C, while moving it to the other direction will decrease C. The capacitive force can also lead to a coupling of polarizations if the force gradients are not parallel to the polarization directions but this only induces a change in polarization axes and doesn't affect the dynamics.

The capacitive force can be written as:
\begin{equation}
\label{capaF}
F_c = C'(\bar{U}+U)^2/2 
\end{equation}
where C' is the derivative of capacitance with respect to x, $\bar{U}$ is the average voltage at the nanowire apex and U is the instantaneous additional voltage. C' is the parameter that controls the breaking of spatial symmetry of the system. In the rest of the text, C' will be considered as negative without loss of generality.

\subsection{Electrical equations}
The voltage at the apex is governed by the Kirchhoff's circuit laws and the way the SEM beam current $I_b$ is divided between the different electrical elements of the system (figure \ref{figelec} b). Primary electrons from the SEM beam can interact with the nanowire. This interaction leads to either an absorbtion of the primary electrons by the nanowire, measured with an electrometer, or an emission of the nanowire electrons into vacuum. The low energy emitted electrons are called secondary electrons emission and are collected by a SED (secondary electron detector). The remaining high energy free electrons are mainly the transmitted electrons, the Auger electrons and the backscattered electrons. The electron current inside the nanowire is controlled by three impedances in parallel : the resistance of the wire, the capacitance of the apex and the "motional impedance" of the mechanical resonator.

\begin{equation}
\label{Idem}
I_b = I_{S}+I_{NW}+I_{H}+I_c+I_m
\end{equation}
where $I_{NW}$ is the current flowing from the nanowire apex to the tungsten tip, $I_{H}$ is the high energy free electron current, $I_c = C\dot{U}$ is the capacitive current, $I_m = C'U\dot{x}$ the motional current \cite{lazarus} and $I_{S}$ is the secondary electron current.

$I_{NW}$ depends on the nanowire resistance $R$ and is given by the Ohm's law :
\begin{equation}
I_{NW} = \frac{U+\bar{U}}{R}
\end{equation}
$I_b$ will be considered as constant, i.e. independent of x and U. $I_{H}$ depends on the thickness of the material and so depends strongly on the relative positions of the nanowire $\bar{x}$ and of the focused beam $x_b$. Moreover the voltage at the apex can influence the amount of transmitted current for instance by slightly deflecting the incident electron beam. For a fixed electron beam position this current is given by :
\begin{equation}
I_{H}(x+\bar{x}-x_b,\bar{U}+U) \approx I_{H}(\bar{x}-x_b,\bar{U})+\frac{\partial I_{H}}{\partial x}(\bar{x}-x_b,\bar{U})x+\frac{\partial I_{H}}{\partial V}(\bar{x}-x_b,\bar{U})U
\end{equation}
As well, the secondary electron emission depends on the volume and the shape of the nanowire, both related to the nanowire position x. It depends also on the apex voltage since the secondary electrons have a low energy and can be recaptured with a positive voltage:
\begin{equation}
\label{ISE}
I_{S}(x+\bar{x}-x_b,\bar{U}+U) \approx I_{S}(\bar{x}-x_b,\bar{U})+\frac{\partial I_{S}}{\partial x}(\bar{x}-x_b,\bar{U})x+\frac{\partial I_{S}}{\partial V}(\bar{x}-x_b,\bar{U})U
\end{equation}
By discarding the constant terms and performing a Fourier transform, the electrical dynamical equation \ref{Idem} becomes :
\begin{equation}
\label{elec1}
\hat{U} = - Z\frac{\partial I}{\partial x}\hat{x} + Z\tilde{I}
\end{equation}
where $\hat{U}$ (respectively $\hat{x}$) is the Fourier transform of U (respectively x), Z is the electrical impedance of the detection circuit, $\frac{\partial I}{\partial x}$ is the electromechanical transduction and $\tilde{I}$ is the total current noise and
\begin{eqnarray}
\label{imped1}
\frac{1}{Z}   & = & \frac{1}{R}+ \frac{\partial I_{S}}{\partial V}+\frac{\partial I_{H}}{\partial V}+i\omega C \\
\frac{\partial I}{\partial x}   & = & \frac{\partial I_{H}}{\partial x}+\frac{\partial I_{S}}{\partial x}+i\omega C'\bar{U}\\
\tilde{I} & = & \tilde{I}_{b}(\omega)+\tilde{I}_{S}(\omega)+\tilde{I}_{H}(\omega)+\frac{\tilde{U}(\omega)}{R}
\label{bruitI}
\end{eqnarray}
where $\omega$ is the angular frequency, $\tilde{I}_{b}$ is the shot noise of the incident electron beam, $\tilde{I}_{S}$ is the secondary electron current noise, $\tilde{I}_{H}$ is the current noise from $I_{H}$ and $\tilde{U}$ is the Johnson noise coming from the resistor. A hidden assumption in this development, is that the change in current is instantaneously related to the change in position and voltage, so that the spatial and voltage derivatives of the current are independent of the frequency. It means for instance that the secondary electron emission process has no delay. We see no reason to doubt the validity of this assumption in the range of frequency and amplitude of vibration considered here. The time scale of the secondary emission is faster than 1 ns, the time scale of the electro-mechanics is usually slower than 1$\mu$s and the amplitude of vibration is smaller than the nanowire diameter or the electron beam width.

\subsection{Effective dynamical equation}

The effective electro-mechanical equation of motion can now be expressed using Eq. \ref{act1}, Eq. \ref{capaF} and Eq. \ref{elec1}:

\begin{equation}
\label{act2}
\chi^{-1}_{e}(\omega)\hat{x} = \tilde{F}_{ba} + \tilde{F_T}
\end{equation}
where $\chi_{e}(\omega)$ is the effective mechanical susceptibility including the backaction capacitive force and $\tilde{F}_{ba}$ is the backaction noise force :
\begin{eqnarray}
\label{sus2}
\chi^{-1}_{e}(\omega) & = & m(\omega_0^2-\omega^2+i\Gamma_0\omega)+C'\bar{U}Z\frac{\partial I}{\partial x}\\
\tilde{F}_{ba} & = & C'\bar{U}Z\tilde{I}
\label{retroF1}
\end{eqnarray}
The last term in Eq. \ref{sus2} is the expression of the complex backaction rigidity $k_{ba}$ due to the interaction of the resonator with the electron beam. This interaction can result in a change in the resonance frequency $\omega_e/2\pi$, a change in the damping $\Gamma_e$ and a resonator effective temperature $T_e$ different from the thermal bath at $T_0$.



\subsection{Effective physical parameters}
The effective temperature $T_{e}$ of the cantilever can be obtained from the equipartition theorem :
\begin{equation}
\frac{1}{2}k_BT_{e} = \frac{1}{2}m\omega_{e}^2<x^2(t)> = \frac{1}{2}m\omega_{e}^2\frac{1}{2\pi}\int_{0}^{\infty}|\hat{x}|^2(\omega)d\omega
\end{equation}
and the effective resonance angular frequency and effective damping are :
\begin{eqnarray}
\omega_{e}^2 & = & \omega_0^2 + \frac{C'\bar{U}}{m} \rm{Re}\left(Z\frac{\partial I}{\partial x}\right)\\
\Gamma_{e} & = &\Gamma_{0} + \frac{C'\bar{U}}{m\omega} \rm{Im}\left(Z\frac{\partial I}{\partial x}\right)
\end{eqnarray}
So
\begin{equation}
\label{Te}
k_BT_{e} \simeq \frac{m\omega_{e}^2}{2\pi}(\tilde{F}^2_T+ \tilde{F}_{ba}^2)\int_{0}^{\infty}\frac{d\omega}{m^2[(\omega^2-\omega_e^2)^2+(\Gamma_e\omega)^2]} \simeq \frac{\tilde{F}^2_T+ \tilde{F}_{ba}^2}{4m\Gamma_e}
\end{equation}
where the following hypothesis has been made : the resonance is sufficiently narrow ($\Gamma_e << \omega_e$) so that the noise can be considered as white, the electrical impedance as constant and the effective susceptibility as a Lorentzian function.

According to the fluctuation dissipation theorem, the power spectrum density of the stochastic thermal force is : \begin{equation}
S_{F_T} = \tilde{F}^2_T = 4 k_BT_0m\Gamma_0
\end{equation}
and from Eq. \ref{bruitI} and \ref{retroF1} the power spectrum density of the backaction noise force is :
\begin{equation}
S_{F_{ba}} = \tilde{F}^2_{ba} = (C'\bar{U}\mid Z\mid)^2(2eI_b+S_{I_{S}}+S_{I_{H}}+\frac{4 k_BT_0}{R})
\end{equation}

The expression of the effective temperature in Eq. \ref{Te} can be rewritten as :
\begin{eqnarray}
\label{TT1}
\frac{T_e}{T_0} & = & \frac{1+\Gamma_0/\Gamma_U+e\tilde{V}/4k_BT_0}{1+\Gamma_0/\Gamma_U+\lambda_C/\lambda_e}\\
\Gamma_U & = & \frac{R(C'\bar{U})^2}{m}\frac{1}{1+(\omega_eRC)^2}\nonumber\\
e\tilde{V} & = & R(2eI_b+S_{I_{S}}+S_{I_{H}})\nonumber\\
\lambda_C & = & -\frac{C}{C'}\nonumber\\
\lambda_e & = & \frac{\bar{U}}{R(\frac{\partial I_{S}}{\partial x}+\frac{\partial I_{H}}{\partial x})} = \frac{\bar{I}_{NW}}{\frac{\partial I_{NW}}{\partial x}}\nonumber
\end{eqnarray}
where we have neglected $\frac{\partial I_{S}}{\partial V}+\frac{\partial I_{H}}{\partial V}$. It appears that the effective temperature depends on: i) $\omega_eRC$ the product of the effective angular frequency by the charge relaxation time; ii) the ratio between the intrinsic damping $\Gamma_0$ and the electrostatic damping $\Gamma_U$; iii) the ratio between the current noise energy $e\tilde{V}$ and the thermal energy. iv) the ratio between the characteristic length of capacitance variation $\lambda_C$ and the characteristic length of free electron variation $\lambda_e$.

\section{Results and discussion}
\subsection{Discussion on the different regimes}

\begin{figure}
\includegraphics[width=12cm]{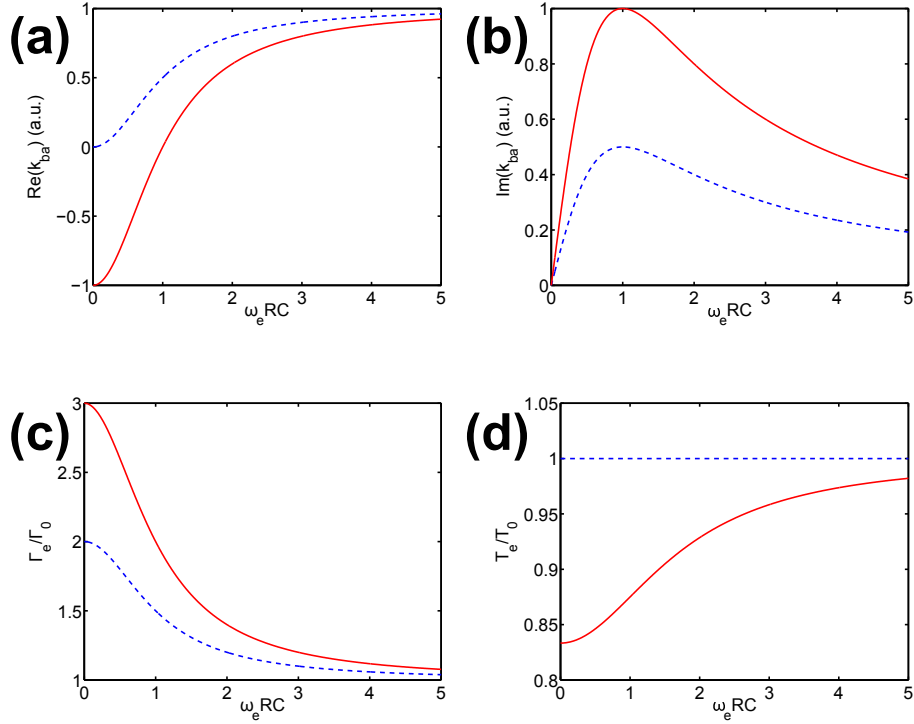}
\caption{\label{figreg} Dynamical regimes of a nanowire interacting with a free electron beam and a purely capacitive force (solid lines) and electrostatic damping regime (dashed lines) as a function of the resonance frequency, a) Evolution of the real part of the backaction rigidity, b) Evolution of the imaginary part of the backaction rigidity, c) Evolution of the damping, d) Evolution of the effective temperature.}
\end{figure}

The complex backaction rigidity in Eq. \ref{sus2} is the product of the electrical impedance and the transduction coefficient. This force strongly depends on the frequency because of the different electrical responsivity regimes and the competition between several transduction mechanisms. In this study, we are mainly concerned by the dynamics around the resonance frequency. This frequency needs to be compared to the electrical angular frequency cut-off  $1/RC$. The effects of the dynamical backaction force are illustrated in Fig. \ref{figreg} by the solid line curves, in the case where $e\tilde{V}/4k_BT_0=0.5$ and $\lambda_e=\lambda_C$, i.e. the transduction strengths of the capacitance and of the free electrons are equal. Although these parameters are not optimal for cooling applications, it illustrates the general trend of the forces originating from the free electron and the electrostatic damping.

At high resonance frequency, the real part of the backaction dominates, mainly inducing a tuning of the effective resonance frequency and leaving the damping unchanged whatever the sign and the amplitude of the free electron transduction. It can be noticed that the imaginary part of the backation rigidity is maximum at the frequency $1/RC$ but it is at a lower frequency that the effect of the backation on the damping and the effective temperature is maximum.

At low resonance frequency, the backaction induces strikingly different effects whether $\lambda_e$ is higher or lower than $\lambda_C$. For $|\lambda_e| << |\lambda_C|$, the situation is close to the case of electrostatic damping\cite{jourdan2007tuning,barois2012ohmic} where the effective damping increases without cooling the resonator and the frequency tuning is negligible. The difference comes from the additional noise from the free electrons. In the absence of this noise (see the curves in dashed line in Fig. \ref{figreg} in the case where the electrostatic damping equals the intrinsic damping), the resonator is submitted to a thermomechanical noise and a Johnson noise from the resistor. When a DC voltage is applied, the resonator damping is higher due to the dissipation in the resistor (Fig. \ref{figreg} c)) but at the same time the Johnson noise generates a noisy capacitive force. This results in an effective temperature of the resonator strictly equal the the bath temperature as the resistor is also in equilibrium with the bath (Fig. \ref{figreg} d)). To the contrary, the presence of the free electron noise, as well as a DC current flowing through the nanowire, generates an additional noisy capacitive force that is not compensated by a change in damping. This additional noise is responsible for an increase of the effective temperature and the higher the free electron current, the higher the temperature.

The case $\lambda_e / \lambda_C<-1$, was studied in Ref.\cite{vincent2007driving} where it was shown that the electron beam can inject energy into the nanocantilever and achieve a self sustained mechanical oscillation regime. As the sign of the free electron transduction can be changed easily by illuminating one side or the other of the nanowire, the sign of the backaction force will also change and thus energy can also be extracted from the resonator. If the energy extraction is higher than the added noise by the free electron current and $\lambda_e / \lambda_C\gtrsim1$, cooling below room temperature is possible as shown in solid line in Fig. \ref{figreg} d). Frequency tuning is possible at low resonance frequency but with an opposite sign as the electrostatic damping tuning at high frequency (Fig. \ref{figreg} a)).
\subsection{Experimental protocol}

The estimation of the expected effective temperature and the determination of the dynamical regimes requires one to know the value of the spatial and voltage derivatives of the current. A rigorous measurement of the backaction force requires to measure the x dependance of both the secondary electron current and the transmitted current, especially since these two current gradients have opposite signs and might cancel each other. For instance, it is possible to modify the yield of secondary electron emission by changing the acceleration voltage. At some specific values of the voltage, there exists a position where the number of absorbed electrons is equal to the number of secondary emitted electrons. In this situation, the secondary electron current and its spatial derivative are not zero whereas no backaction should take place as the voltage at the apex is zero.  Moreover, it is necessary to sweep only one variable while maintaining the others constant. However, a change of the voltage will usually induce a change in current but also a change in position due to the change of the electrostatic deflection force. Therefore in the following, an experimental protocol will be defined in order to obtain a reliable estimate of the experimental parameters from our previous experiments\cite{vincent2007driving,ayari_self_oscillations,PhysRevBself}.

\subsubsection{DC current and voltage}
The experiments were performed in a Hitachi S800  or a Orsay Physics e-CLIPSE. The acceleration voltage was typically in the tens of kV and the beam current was about 100 pA (measured with a faraday cup). By focusing the electron beam at the apex of a SiC nanowire electrically connected to a Keithley 6517 electrometer, we measured the DC current flowing through the nanowire. Measuring I$_S$ and I$_H$ is rather difficult. A measurement of I$_S$ can be done with a secondary electron detector but usually the detector collects a limited part of the solid angle of emission and can give angle-dependent, spurious results if the sample surface is not flat. I$_H$ can be measured if an additional Faraday cup is properly positioned below the sample. A measurement of I$_{NW}$ is simpler and will give in a single measurement the sum of I$_S$ and I$_H$. By moving the electron beam spot along the nanowire diameter, we observe a maximum current at the position where the beam crosses the thicker part of the wire. In a range of 10 nm around the position of maximum current, this maximum doesn't change significantly (i.e. lower than $5 \%$ of variation). The maximum current $I_{NW}$ flowing through a SiC nanowire is lower than 50 pA. It varies from sample to sample and can be as low as 2 pA. From the sign of the current, we were able to determine that the electrons flow from the tungsten tip to the apex, i.e. in the case of SiC, more electrons are emitted from the nanowire apex than absorbed.

The voltage is obtained indirectly by estimating the nanowire resistance by the method detailed in ref. \cite{ayari_self_oscillations,PhysRevBself}. The typical resistance is about 1 G$\Omega$ and in some exceptional cases it can reach up to 1 T$\Omega$ but we never performed extensive experiments on such highly resistive samples. The typical DC voltage $\bar{U}$ is lower than 1V.


\subsubsection{Voltage derivative of the current}
The voltage derivative of the current can be measured by placing the electron beam in the region of the nanowire where the spatial derivative of the current is zero. As already mentioned above, by slowly moving the electron beam in the region where the beam crosses the thicker part of the wire, $I_{NW}$ is constant in a range of at least 10 nm. So the variation of $I_{NW}$ in this region is :
\begin{equation}
\delta I_{NW} = \frac{\delta U}{R} =  (\frac{\partial I_{S}}{\partial x}+\frac{\partial I_{H}}{\partial x})\delta x+(\frac{\partial I_{S}}{\partial V}+\frac{\partial I_{H}}{\partial V})\delta U = 0
\end{equation}
as $I_b$ is constant and $I_c$ ant $I_m$ are zero for slow displacements (i.e. $\omega = 0$). So $\delta U = 0$ and $\frac{\partial I_{S}}{\partial x}+\frac{\partial I_{H}}{\partial x}=0$ in this region. Then, the position of the beam is fixed and a DC voltage is applied to the tungsten tip. At this position a change of $I_{NW}$ with the applied voltage gives the voltage derivative of the sum of $I_{S}$ and $I_{H}$ as long as the nanowire deflection due to the capacitive force doesn't exceed 10 nm. This hypothesis can be easily checked by taking SEM images of the entire nanowire first at zero voltage and various beam currents in order to determine the maximum beam current allowed that does not induce electrostatic bending; then for several sample voltages to measure its bending. The deflection depends on the square law of the voltage, so the highest acceptable voltage can be easily deduced. In our experiments, $\frac{\partial I_{S}}{\partial V}+\frac{\partial I_{H}}{\partial V}$ never exceeds 1 pA for a voltage change of 1V. This gives an effective differential resistance above 1 T$\Omega$. We deduce from this estimation that in the expression of the impedance Z in Eq. \ref{imped1} the voltage derivative of the free electron currents can be safely ignored compared to the sample resistance.

In the regions of the nanowire where the spatial derivatives of the currents are not zero, the voltage derivative cannot be measured independently. However, it is reasonable to consider that $\frac{\partial I_{S}}{\partial V}+\frac{\partial I_{H}}{\partial V}$ is smaller at the edge of the nanowire than in the middle because $I_{NW}$ and thus $\bar{U}$ are smaller.

\subsubsection{Spatial derivative of the current}
As the typical current flowing through the nanowire is of the order of several tens of pA and the spatial range where the current changes significantly at the edge is about 10 nm, a rough estimate of the spatial derivative of the current is 1pA/nm. This value is an order of magnitude higher than the spatial derivative of the field emission current in ref. \cite{PhysRevBself} for the same DC current.

\subsubsection{Other physical parameters}

The method to estimate C and C' has been explained elsewhere\cite{PhysRevBself}. The capacitance is usually around 1 fF which gives an electrical frequency cut-off between 1 kHz and 1 MHz. So, in our experiments $\omega_eRC$ can range from $\sim0.1$ (Doppler regime) to above 10 (resolved side band regime).

$|C'|$ is lower or of the order of 1 pF/m. $\lambda_C$ is approximately of the order of the distance between the tip and the counter electrode and is independent of the beam current. $\lambda_e$ strongly depends on the electron beam position and is rather independent of the beam current as $\bar{U}$ and the spatial derivative of the current are both proportional to the current.

The ratio $e\tilde{V}/4k_BT_0$ dominates at high beam current and tends to increase the effective temperature. An estimation of the total current noise will be rather crude as it involves many different processes not easily accessible experimentally. The SEM beam current will be supposed to be shot noise limited. The current noise measured on the SED was white in a frequency range around the resonance frequency and proportional to the collected secondary electrons with a Fano factor between 2 and 3 (i.e. 2 or 3 times noisier than a pure shot noise). Concerning $\tilde{I}_H$, it can be reasonably guessed that it will also be super Poissonian with a similar Fano factor and partially correlated with the electron beam shot noise.

\subsection{Estimation of the effective temperature}

At this stage, all the required parameters for the model have been reasonably estimated. Now, analytical and numerical calculations can be performed in order to illustrate the strength of the electron beam interaction in a typical example. We will consider a SiC nanowire with a Young's modulus of 500 Gpa, a density of 3210 $kg/m^3$, a quality factor of 10 000, a length of 30 $\mu$m and a radius of 30 nm. Thus, the mass is equal to 8.2 $\cdot$10$^{-17}$ kg and the resonance frequency is equal to 128 kHz. The capacitance comes from a metallic plate parallel to the nanowire at a distance of 10 $\mu$m. So an electrostatic calculation gives C = 0.26 fF and C' = 2.03 pF/m at 0 V. The nanowire resistance is 3 G$\Omega$. The total beam current is 100 pA with a width of 10 nm and Gaussian profile. The yield of electron absorption is chosen to be 10 $\%$ in the center where the thickness is higher, and is proportional to the nanowire thickness. On the edge of the nanowire, for I$_{NW}$=6 pA, its spatial derivative equals 0.3 pA/nm.  As a safe estimate, we will consider that the total current noise is ten times the shot noise of the total incident beam current. The true noise value is probably lower, so the effects will be stronger experimentally than in our calculations.

With this choice of parameters, the calculation of the ratios involved in Eq. \ref{TT1} gives: i) $\omega_eRC \approx 0.62 $;  ii) $\Gamma_0/\Gamma_U \approx 1700$, the electrostatic damping is negligible; iii) $e\tilde{V}/4k_BT_0 \approx 1.8$, the current noise is of the order of the thermal noise but is negligible compared to the thermomechanical noise (1700 $>>$ 1.8) because the capacitive transduction is too low. So the electron beam doesn't induce additional noise; iv) $\lambda_C/\lambda_e \approx 5000$, the system is in the strong free electron transduction regime inducing an increase of the damping. Then, the effective temperature according to Eq. \ref{TT1} is 4 times lower than the bath temperature at this position, T$_e$ = 75 K for a bath at 300 K.

\subsection{Numerical simulations}

\begin{figure}
\includegraphics[width=8cm]{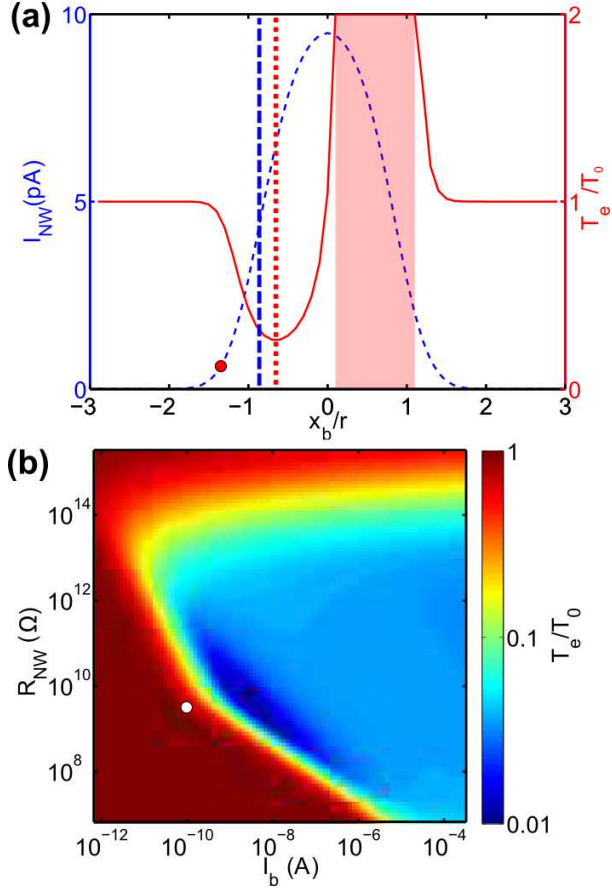}
\caption{\label{figsim} a) Current flowing through the nanowire (doted line) and ratio of the effective temperature on the bath temperature (solid line) as a function of the position of the electron beam for an incident beam current of 100 pA. The red circle indicates the chosen position of the map in the figure b. The shaded area represents the self-oscillation regime. The vertical dashed line indicates the beam position where the spatial derivative of the current is maximal. The vertical dotted line indicates the beam position where the cooling is maximal. b) Evolution of the effective temperature as a function of the resistance and beam current. The white circle indicates the parameters used for the profile in the figure a.}
\end{figure}

We performed numerical simulations with a self-consistant determination of the apex voltage,  position, capacitance and its spatial derivative as a function of the beam position. Fig.\ref{figsim} a) represents the profile of the current absorbed by the nanowire for a Gaussian incident beam at different positions perpendicular to the length of the wire. On one edge of the nanowire, the electron beam reduces the damping leading to self-oscillation and the effective temperature diverges as observed experimentally. On the other edge, the effective temperature is reduced and the value of $T_{eff}$ is in good agreement with the analytical calculation. This calculation demonstrates that for parameters compatible with typical experiments, the cooling effect mediated by a capacitive force is important. This is the main result of this article.

With this model, it is possible to identify the influence of each physical parameters. For instance in Fig. \ref{figsim} b), we swept the nanowire resistance and beam current to look for the optimal parameters for a given position of the beam. In this conditions, $\lambda_C$ and $\lambda_e$ are roughly constant and the optimum is a trade off between a high voltage for a high capacitive transduction and a low current to limit the effective heating from the electrical noise (e$\tilde{V}$ term). We obtained that a nanowire with a resistance of 4 G$\Omega$ and a beam current of 3 nA can be cooled down by two orders of magnitude (3 K).

The most important parameter to optimise is the position of the electron beam. In Fig. \ref{figsim} a), it can be seen that contrary to what could be intuitively guessed, the optimal cooling is obtained for a beam position different than the one where the spatial derivative of the current is maximal. A more careful analysis shows that optimizing the beam position requires to maximize $\lambda_e$. For an electron beam with a Gaussian profile at fixed I$_{NW}$, it turns out that $\lambda_e$ can be increased indefinitely. An increase of the difference between the beam position and the nanowire center position and a concomitant increase of I$_b$ in order to maintain the absorbed current I$_{NW}$ constant, will lead to a higher value of $\lambda_e$. In other word, the further away the beam, the higher the cooling. However, experimentally, in most SEMs, increasing the current is done by increasing the acceleration voltage or increasing the spot size. Increasing the voltage might reduce the number of absorbed electrons and increasing the spot size will strongly decrease $\lambda_e$. Indeed, the reason why $\lambda_e$ is an order of magnitude better than what can be obtained in field emission NEMS\cite{ayari_self_oscillations,PhysRevBself} comes essentially from the SEM high resolution. Optimizing the beam current and position is strongly dependent on the electron microscope performances and material properties. In our SEMs, $\lambda_e$ is typically around 10 nm and can probably be improved by one order of magnitude. Then, it would be comparable to the characteristic length obtained at low temperature in quantum point contact (QPC) mechanical sensors\cite{flowers2007intrinsic,poggio2008off} or wide band scanning tunneling microscopes (STM)\cite{kemiktarak2007radio} ranging from 2.5 nm down to 0.1 nm. However, using a SEM is as flexible as the off-board detection of ref. \cite{poggio2008off} and its main advantage is that it doesn't require a cryogenic environment. Reaching ground state cooling seems nevertheless rather hard with this technique and would definitely require to work with a transmission electron microscope.

\subsection{Motion sensing}

If the electron beam is used for position sensing instead of cooling, the optimal conditions are different. The expression of the secondary electron current from Eq. \ref{ISE} reveals that using this current as a position sensor like for instance in ref. \cite{PhysRevB.63.033402,vincent2007driving} might be problematic as it mixes mechanical and electrical information. $I_{S}$ depends not only on x but also on U. x and U are two independent dynamical variables related by the electromechanical equations. By performing a Fourier transform of Eq. \ref{ISE}, leaving aside the noise terms and using Eq. \ref{elec1}, the relation between the secondary current and the nanowire position becomes :
\begin{equation}
\label{Isense}
\hat{I}_{S} \approx \frac{\partial I_{S}}{\partial x}\hat{x}-\frac{\partial I_{S}}{\partial V}Z\frac{\partial I}{\partial x}\hat{x}
\end{equation}
This relation shows that $\hat{I}_{S}$ and $\hat{x}$ are proportional but the coefficient of proportionality might depend on the frequency of interest since Z and $\frac{\partial I}{\partial x}$ are frequency dependent. For instance, calibrating the displacement by performing a line mode scan (i.e. slowly sweeping the electron beam along a line perpendicular to the nanowire) while recording the secondary emission current  might be incorrect when studying the mechanical response at high frequency. Moreover, the second term in Eq. \ref{Isense} indicates that in the middle of the wire where $\frac{\partial I_{S}}{\partial x}$ is expected to cancel out, measuring the displacement is in principle still possible as the electromechanical transduction $\frac{\partial I}{\partial x}$ is non zero and dominated by the C' term. Such position might be interesting, if one wants to reduce the backaction force of the beam. For our sample, this term is negligible as $|Z\partial I_{S}/\partial V|<< 1$ and plays a role only for highly resistive samples $R \gtrsim 1T\Omega$ or for a higher beam current.

Now, if we take into account the noise terms in Eq. \ref{ISE}, \ref{elec1}, \ref{act2} and \ref{retroF1}, the Fourier transform of the mechanical displacement is
\begin{equation}
\label{Isense2}
\hat{I}_{S} = \tilde{I}_{S}(\omega)+\frac{\partial I_{S}}{\partial x}\chi_{e}(\omega)C'\bar{U}Z\tilde{I} + \frac{\partial I_{S}}{\partial x}\chi_{e}(\omega) \tilde{F_T}
\end{equation}
where the voltage derivative of $I_{S}$ has been neglected. If we follow the same line of reasoning as in optomechanics \cite{braginsky1995quantum,aspelmeyer2014cavity}, the first term on the right hand side of Eq. \ref{Isense2} is the imprecision noise current and the second term is the backaction noise force. The third term is the quantity we need to measure. These terms are usually compared when referred back to the input i.e. by dividing them by the transduction $\frac{\partial I_{S}}{\partial x}$.
For the previous typical sample we used to calculate the effective temperature in Fig. \ref{figsim}, the thermomechanical motion at the resonance is 1 nm/$\sqrt{Hz}$ without the electron beam and 250 pm/$\sqrt{Hz}$ when the beam is at the optimal cooling position.
The power spectrum density of the imprecision displacement noise can be expressed as :
\begin{equation}
S_x^{imp} = \frac{S_{I_{S}}}{(\frac{\partial I_{S}}{\partial x})^2} \simeq \frac{2e\alpha\lambda_S^2}{I_b\eta}
\end{equation}
where $\lambda_S$ is the the characteristic length of secondary electron current variation and will be considered as close to $\lambda_e$, $\alpha$ is the Fano factor expressing the excess noise compared to the shot noise, $\eta$ is the yield of conversion of incident electron into secondary electrons. We will  use the same conservative estimate for the noise as previously so $\alpha = 10$ and fix the yield at 10 $\%$ as before. As expected the imprecision noise decreases when the incident beam current increases. For our typical example, the imprecision displacement noise is lower than 12 pm/$\sqrt{Hz}$.

The power spectrum density of the backaction displacement noise is
\begin{equation}
S_x^{ba}(\omega_e) \simeq \left(\frac{C'R^2\eta}{m\omega_e\Gamma_e}\right)^2 2e\alpha I_b^3
\end{equation}
where we made the supplementary approximation $\mid Z\mid=R$. For our typical example, the backaction displacement noise is 16 pm/$\sqrt{Hz}$ very close to the imprecision displacement noise and negligible compared to the thermomechanical noise. So this configuration is close to the optimal beam current value as increasing further the current would reduce the imprecision noise but will increase the backaction noise and increase the damping making detection more difficult. A rigorous analytical determination of the optimal current would require to estimate the cross correlation between the noise current and the backaction noise and will depend on the frequency of interest as in optomechanics\cite{PhysRevB.70.245306}(for example, at the resonance $\Gamma_e$ is in our case proportional to $I_b^2$). This optimum can be obtained by minimizing the ratio of the sum of the power spectrum density of the imprecision displacement noise, the backaction displacement noise and their cross correlation, by the power spectrum density of the thermomechanical noise. The total additional noise in our typical example is of the order of several tens of pm/$\sqrt{Hz}$. It is still far from the best detection methods involving electrons in vacuum such as quantum point contacts\cite{flowers2007intrinsic} and STM\cite{kemiktarak2007radio} which have a sensitivity in the fm/$\sqrt{Hz}$. However, we consider our estimation as very conservative and despite this, the noise is low enough to measure the thermomechanical noise with at least 20 dB of signal to noise ratio. Our total additional noise is comparable to calculations for the optimal condition in current mixing in single electron transistor double clamped carbon nanotubes\cite{PhysRevB.95.035410} at low temperature and probably close to the performances of an off-board QPC at low temperature\cite{poggio2008off}.

\section{Conclusions}

We showed analytically that the capacitive interaction between a scanning electron microscope electron beam and a SiC nanowire mechanical resonator can have a strong influence on the effective temperature of the resonator. Our numerical simulations with parameters based on experimental conditions indicate that lowering the temperature below 4K can be easily obtained. Interestingly, the beam position where maximum cooling efficiency takes place is dependent on the total beam current and is not obtained where the spatial variation of the current is maximum.
The importance of measuring the absorbed electron current on the nanowire instead of the secondary electron current has been highlighted to properly measure the backaction force from the beam. We demonstrated that secondary electrons can be used as position sensors if the induced voltage is not too important. A beam current of 100 pA is close to the optimal condition where the backcation noise force and the imprecision noise are minimized and have a value of the order of 10 pm/$\sqrt{Hz}$. Some improvements are still necessary to compete with other techniques using vacuum electrons but a careful choice of material parameters and geometry as well as the use of a transmission electron microscope may strongly improve future results.



%
%

%


%

\end{document}